\documentstyle[graphicx,psfig,rotating]{mn}
\newcommand{\errtwa}[3]{#1^{+#2}_{-#3}}
\newcommand{\newblock}{}
\newcommand\aproxgt{\mathrel{%
      \rlap{\raise 0.511ex \hbox{$>$}}{\lower 0.511ex \hbox{$\sim$}}}}
\newcommand\aproxlt{\mathrel{%
      \rlap{\raise 0.511ex \hbox{$<$}}{\lower 0.511ex \hbox{$\sim$}}}}
\newcommand\eight{{X~1822$-$371}}
\newcommand\rxte{\textsl{RXTE}}
\newcommand\pca{\textsl{PCA}}
\newcommand\hexte{\textsl{HEXTE}}
\newcommand\asca{\textsl{ASCA}}
\newcommand\sis{\textsl{SIS}}
\newcommand\gis{\textsl{GIS}}
\newcommand\astroe{\textsl{Astro-E}}
\newcommand\exosat{\textsl{EXOSAT}}
\newcommand\ginga{\textsl{Ginga}}
\newcommand\einstein{\textsl{Einstein}}
\newcommand\beppo{\textsl{BeppoSAX}}
\newcommand\xmm{\textsl{XMM-Newton}}
\newcommand\axaf{\textsl{Chandra}}
\newcommand\msun{{\rm M_\odot}}

\begin{document} 
\title[The Flared Disc Project: RXTE and ASCA observations of \eight]{The
  Flared Disc Project: RXTE and ASCA observations of \eight}
\author[S. Heinz \& M. A. Nowak]{S.~Heinz \& M.~A.~Nowak\\
  JILA, University of Colorado, Campus Box 440, Boulder, CO 80309-0440}
\maketitle
\begin{abstract}
  We present archival \textsl {Rossi X-ray Timing Explorer} (\rxte) and
  simultaneous \textsl{Advanced Satellite for Cosmology and Astrophysics}
  (\asca) data of the eclipsing low mass X-ray binary (LMXB) \eight.  Our
  spectral analysis shows that a variety of simple models can fit the
  spectra relatively well.  Of these models, we explore two in detail
  through phase resolved fits. These two models represent the case of a
  very optically thick and a very optically thin corona.  While systematic
  residuals remain at high energies, the overall spectral shape is
  well-approximated.  The same two basic models are fit to the X-ray light
  curve, which shows sinusoidal modulations interpreted as absorption by an
  opaque disc rim of varying height.  The geometry we infer from these fits
  is consistent with previous studies: the disc rim reaches out to the tidal
  truncation radius, while the radius of the corona (approximated as
  spherical) is very close to the circularization radius.  Timing analysis
  of the \rxte\ data shows a time lag from hard to soft consistent with the
  coronal size inferred from the fits. Neither the spectra nor the light
  curve fits allow us to rule out either model, leaving a key ingredient of
  the \eight\ puzzle unsolved.  Furthermore, while previous studies were
  consistent with the central object being a $1.4~\msun$ neutron star,
  which has been adopted as the best guess scenario for this system, our
  light curve fits show that a white dwarf or black hole primary can work
  just as well. Based on previously published estimates of the orbital
  evolution of \eight, however, we suggest that this system contains either
  a neutron star or a low mass ($\aproxlt 2.5~\msun$) black hole and is in
  a transitional state of duration shortward of $10^{7}$ years.
\end{abstract}
\section{Introduction}\label{sec:intro}

\eight\ is a low mass X-ray binary and is the prototypical `Accretion Disc
Corona' (ADC) source \cite{white:82a}. The 5.57\,hr orbital period of
\eight\ \cite{white:81a} exhibits a quasi-sinusoidal variation in both the
X-ray and optical (see also Mason \& C\'ordova \nocite{mason:82a} 1982;
Hellier \& Mason \nocite{hellier:89a} 1989).  In addition, there is an
approximately 20\,min long dip associated with partial obscuration of the
X-ray source by the secondary mass-donating star (presumed to be filling
its Roche lobe; see White et al.  \nocite{white:81a} 1981, White \& Holt
\nocite{white:82a} 1982, Mason \& C\'ordova \nocite{mason:82a} 1982,
Hellier \& Mason \nocite{hellier:89a} 1989, and Hellier et al.
\nocite{hellier:92a} 1992).  Thus \eight\ is believed to be a near edge on
source; however, since the X-ray dip--- henceforth defined to be at zero
orbital phase--- is only partial, the X-rays are presumed to emanate from a
very extended corona with a radius of the order of $3 \times 10^{10}$\,cm.

Assuming a distance of 2\,kpc \cite{mason:82a}, the observed X-ray flux
corresponds to an isotropic luminosity of $L_{\rm iso} \sim 10^{36}~{\rm
  ergs~s^{-1}}$. The central X-ray source is obscured, however, and we only
observe X-rays scattered into our line of sight.  The intrinsic X-ray
luminosity is therefore undoubtedly greater, perhaps substantially so (see
\S\ref{sec:discuss}).  The properties and origin of this scattering corona
are largely unknown.  White \& Holt \shortcite{white:82a} postulated a
corona, possibly optically thick, driven by a photoionizing radiation flux
near the Eddington luminosity.  In their model, a near Eddington luminosity
is required to achieve the large scale height of the corona.  Frank et al.
\shortcite{frank:87a} postulated that the corona was due to interaction of
the incoming accretion stream with the disc at the circularization radius.
Viscous dissipation at radii within the circularization radius would in
part lead to the vertical extent of the corona.  

The sinusoidal modulation is associated with obscuration by material with a
vertical extent of order $10^{10}$\,cm \cite{white:82a,hellier:89a}.  Frank
et al. \shortcite{frank:87a} postulated that this obscuring material is
also located at the disc circularization radius. However, the fact that the
optical lightcurve also shows a sinusoidal modulation and a dip with
approximately twice the duration of that in the X-ray suggests that this
obscuring rim is actually at twice this radius, i.e., closer to the disc
tidal truncation radius (Hellier \& Mason \nocite{hellier:89a} 1989; see
also \S\ref{sec:lcfit} below). Again, this rim is associated with the
interaction of the accretion stream with the accretion disc
\cite{armitage:96a,armitage:98a}.

Perhaps the most confusing aspect of \eight\ has been its spectra.  White
et al. \shortcite{white:81a} fit \einstein\ spectra with a flat,
exponentially cutoff power law (photon index, $\Gamma \sim -1$, cutoff
energy $\approx 17$\,keV).  Furthermore, they required a broad (4\,keV wide)
Fe line component with equivalent width $\sim 1$\,keV and a soft excess
that they attributed to either a 0.25\,keV blackbody or possibly an Fe L
complex.  Hellier \& Mason \shortcite{hellier:89a}, on the other hand, fit
\exosat\ spectra with a flat power law ($\Gamma \approx -0.8$), an $\approx
2$\,keV blackbody, and an Fe K$\alpha$ line with 270 eV equivalent width.
The flux of the blackbody had an implied emitting area consistent with
1/400 of the surface area of a neutron star.  Hellier \& Mason
\shortcite{hellier:89a} therefore postulated that this emission is indeed
from a neutron star surface and is scattered into our line of sight by a
very optically thin corona.

Several years later, Hellier et al. \shortcite{hellier:92a} considered
\ginga\ spectra of \eight, and they attempted to fit the same model as for
the \exosat\ data.  Although such a model was the best simple fit that they
could achieve, the fits were not adequate and yielded a reduced $\chi^2
\sim 10$.  Furthermore, Hellier et al. noted that the dip at zero orbital
phase was fractionally larger at higher energies.  They interpreted this
fact as indicating that the corona was in fact optically thick.  Hellier et
al.  \shortcite{hellier:90a} claim that over the orbital timescales, the Fe
K$\alpha/\beta$ band shows greater variability than the continuum bands.
More recently, Parmar et al. \shortcite{parmar:00a} (hereafter P00)
examined \beppo\ and \textsl{Advanced Satellite for Cosmology and
  Astrophysics} (\asca) data of \eight. (We consider some of the same
\asca\ data in \S\ref{sec:data} below.)  They consider a model comprised of
Comptonization of a Wien tail ($kT_{\rm W} \approx 150$\,eV) in a very
optically thick, low temperature corona ($\tau_{\rm es} \approx 23$,
$kT_{\rm e} \approx 7$\,keV). Also required in these fits are a blackbody
with $kT \approx 1.8$\,keV and an Fe K$\alpha$ line with equivalent width
65--150\,eV. This latter feature possibly could be in reality two lines (Fe
K$\alpha/\beta$), not adequately resolved from one another.  P00 found no
evidence for either an Fe K-edge or O K-edge; however, they claim a
detection of a 1.3--1.4\,keV edge with $\tau\approx 0.1$--$0.3$ that they
associate with K-edges of Ne X and neutral Mg, or the L-edges of moderately
ionized Fe.  (See our discussion of \S\ref{sec:discuss}, however.)

The question of whether \eight\ contains a neutron star, black hole, or
even a white dwarf primary also is yet unresolved.  The lightcurves have
been fitted with a model with a $1.4~\msun$ primary \cite{hellier:89a};
however, as we discuss in \S\ref{sec:lcfit} these fits do not uniquely
determine the primary mass. To date, no variability has been detected that
would uniquely point to a neutron star primary (see Hellier et al.
\nocite{hellier:90a} 1990 for variability analysis of the \ginga\ data).
Previous analyses only detected variability on the orbital time scales.

Facing a new generation of X-ray satellites, we decided to revisit \eight\ 
with the help of predominantly unpublished simultaneous archival
\textsl{Rossi X-ray Timing Explorer} (\rxte) and \asca\ data.  Our goal is
to assess the evidence gathered so far and to point out what we believe to
be future avenues for cutting edge X-ray spectroscopy, as will be provided
by the \textsl{X-ray Multiple Mirror-Newton} telescope ({\xmm}) and \axaf.
First, we discuss the orbital ephemeris (\S\ref{sec:orbit}) and the gross
spectral variations over the orbit (\S\ref{sec:spectra}).  We then consider
two separate spectral models which can be considered as broadly
representing an `optically thick corona' and an `optically thin corona'
(\S\ref{sec:spectra}). Using the \rxte\ lightcurves, we consider fits in
multiple energy bands (\S\ref{sec:lcfit}).  Here we consider both optically
thick and optically thin coronae, and furthermore we consider white dwarf,
neutron star, and low mass black hole primaries.  The high frequency
($10^{-3}$--$0.3$\,Hz) variability of the \rxte\ data is then considered
(\S\ref{sec:var}).  We then discuss the implications of these analyses
(\S\ref{sec:discuss}) and present our conclusions
(\S\ref{sec:conclusions}).

\section{Data Analysis}\label{sec:data}
In this paper we analyze a set of simultaneous \rxte\ and \asca\ 
observations of \eight\ taken in 1996 and a separate set of \asca\ 
observations from 1993, as summarized in Table \ref{tab:instruments}.
Details of the data reduction procedure are summarized in the Appendix. For
\rxte\ we only consider data from the \textsl{Proportional Counter Array}
(\pca), while for \asca\ we consider data from both the \textsl{Solid State
  Imaging Spectrometers} (\sis) and the \textsl{Gas Imaging Spectrometers}
(\gis).

\begin{table}
\caption{Observation Log. \protect{\label{tab:instruments}}}
{\small \begin{tabular}{lccrc}
Satellite & Instr. & OBS ID & Start Date & Exp. Time \\
\hline
\noalign{\vspace*{0.7mm}}
RXTE&PCA&10115-01&26 Sep. 96&21.6 ksec\\
ASCA&SIS,GIS&44015000&26 Sep. 96&28.5 ksec\\
ASCA&SIS,GIS&40019000&7 Oct. 93&38.5 ksec
\end{tabular}}
\end{table}

We use the simultaneous \rxte/\asca\ data to perform `global', broad-band
fits, and not for detailed line spectroscopy.  In what follows, we
therefore have combined the two separate \sis\ spectra into one spectrum
and likewise we have combined the two separate \gis\ spectra into a single
spectrum. As we use the 1993 \asca\ spectra for detailed line modelling, we
do not combine the separate \sis\ spectra for those observations.

\subsection{Orbital Evolution}\label{sec:orbit}
Updated parameters on ephemeris and period changes have most recently been
provided by P00, including all three data sets analyzed in this paper.  To
confirm the measurements by P00, we fitted the eclipse with a linear function
attenuated by a Gaussian.  Our fits agree with those of P00 to within the
error bars (indicated by parentheses below).  P00 et al. give a quadratic
ephemeris for the eclipse midpoint as a function of orbital number, $N$, of
\begin{eqnarray}
T_{\rm ecl} & = & 2445615.30964(15) + 0.232108785(50){\rm N} 
\cr
 & & + 2.06(23)\times10^{-11}{\rm N}^2 ~~.
\end{eqnarray}
This implies a period derivative of
\begin{equation}
  \frac{dP}{dt} = 1.78(20)\times10^{-10} ~~.
\end{equation}
(See also Hellier et al. \nocite{hellier:90a} 1990.)  Thus $\dot P^{-1} P
\approx 3.6\times10^{6}$\,years.

We can now use this value to estimate the mass transfer rate in the system.
We assume that a fraction $1-f$ of the mass lost from the secondary is
transferred to the primary, while the remaining fraction $f$ is lost from
the system at the location of the primary (e.g., in the form of a central
disc wind or jet): $\dot{M_{1}}=-(1-f)\dot{M_{2}}$.  The angular momentum
loss is then $\dot{J}/J = f \dot{M_{2}}/M_{2} q^2/(1+q)$ .  If we define
the usual mass ratio $q \equiv {M_{2}}/{M_{1}}$, where $M_{1}$ and $M_{2}$
are the masses of the primary and the secondary respectively, we can write
the mass accretion rate onto the primary as
\begin{equation}
  \dot{M_{1}} = M_{1} \frac{q}{3
  (1-q)}\frac{\dot{P}}{P}\left(1-f\right)\left[1 + 
  \frac{2}{3}\frac{fq}{1 - q^2}\right]^{-1} ~~.
\label{eq:m1dot}
\end{equation}
For $q = 0.2$ and $M_{1} = 1.4~\msun$ (see \S\ref{sec:lcfit}) this gives
\begin{eqnarray}
  \dot{M_{1}} &=& 2.1 \times 10^{18} \left(1 - f\right)\left[1 + 0.14
  f\right]^{-1} {\rm g \, s^{-1}} 
\cr
  &=& 3.3 \times 10^{-8} \left(1 - f\right)\left[1 + 0.14
  f\right]^{-1} {\rm \msun \, yr^{-1}} ~~.
\label{eq:mdot}
\end{eqnarray}
Note that the Eddington accretion rate for a $1~\msun$ compact object is
$\dot M_{\rm Edd} \sim 1.5\times10^{18}\left(0.1/\eta\right){\rm
  g\,sec^{-1}}$, where $\eta$ is the radiative efficiency.  The second term
in square brackets of eq.\,(\ref{eq:m1dot}) is small for $q < 0.7$.  If, on
the other hand, most of the mass loss occurs in the form of a wind from the
secondary, the estimate for $\dot{M_{1}}$ is much less well-constrained.
Since \eight\ is a LMXB, however, it is somewhat unlikely that the
low-mass companion will have a very strong wind. 

The above mass loss rate implies a change of the Roche lobe radius of the
secondary, $R_2$, on comparably short time scales as the orbital period
evolution.  Specifically, one can show that given the above assumptions
\begin{equation} \frac{\dot R_2}{R_2} = -2 \left [ \frac{5}{6} - q +
\frac{f q}{3 (1+q)} \right ] \frac{\dot M_2}{M_2} ~~.  \label{eq:rdot}
\end{equation} That is, for $q \approx 0.2$, the secondary's Roche lobe
radius is expanding, even for large mass loss via a wind, and therefore we
expect strong mass transfer to be a short-lived phenomenon in this
system. This conclusion is unaltered even if the (non-magnetic) wind mass
loss occurs at radii as large as the disc circularization radius.  If
instead we postulate conservative mass transfer but an unspecified source
of angular momentum loss (perhaps from magnetic braking and gravitational
radiation) sufficiently large to lead to $\dot R_2 = 0$, this implies an
even larger mass transfer rate than discussed above ($\dot M_1 \approx
10^{19}~{\rm g~s^{-1}}$ for a $1.4~\msun$ primary), and also implies a
characteristic angular momentum loss time scale of order $10^7~{\rm
  years}$.  We return to these considerations in \S\ref{sec:discuss}.

\subsection{Spectral Analysis}\label{sec:spectra}
Given the uncertainties in the physical conditions in this system we felt it
appropriate to keep spectral fitting to a phenomenological level. The fits
we will present in the following are meant to represent just the essential
features of the two physical cases outlined in the introduction: optically
thick and optically thin geometries.

\begin{figure}
  \psfig{figure=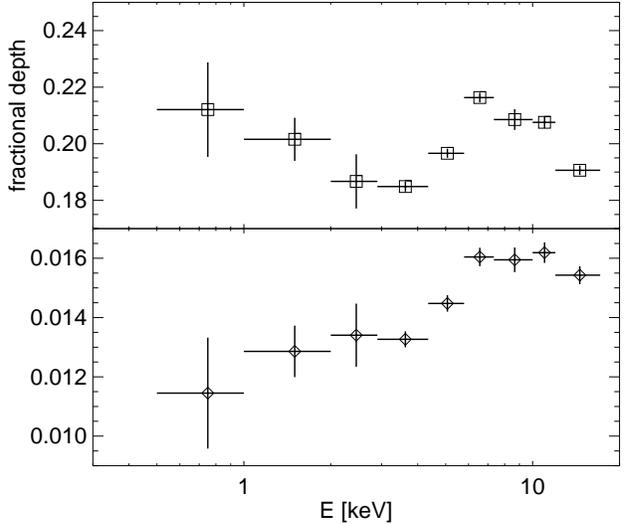,width=\columnwidth}
  \caption{Fractional depth of sinusoidal lightcurve variation
    (top panel, squares). The bottom panel (diamonds) show the product of
    amplitude times width of the Gaussian eclipse (both width and amplitude
    increase with energy).}
  \label{fig:depths}
\end{figure}

\exosat\ observations show an energy dependence of the modulation depth
both for the sinusoidal component due to the accretion disc rim and the
quasi-Gaussian eclipse due to the companion \cite{hellier:89a}. Such a
dependence might indicate a temperature stratification in the coronal
region. We have tested the simultaneous \rxte\ and \asca\ observations for
such a trend by fitting a series of harmonics and a Gaussian to the X-ray
lightcurves of the three instruments (\pca, \sis, and \gis). After
inspection for convergence we truncated the fitting at the third harmonic.
The fits confirm the trends reported by Hellier \& Mason
\shortcite{hellier:89a}.  The fractional depth of the lowest order harmonic
decreases with energy between 0.6 keV and 3 keV, and then rises again near
the Fe K$\alpha$/K$\beta$ region. The variation is small (of order 10\%),
but it is statistically significant. Both the fractional depth of the
Gaussian and its width increase with energy. The product of the two
(proportional to the fraction of the emission blocked by the secondary)
actually varies significantly with energy, roughly by 50\%, which we plot
in Fig.\ref{fig:depths}.

Due to the energy dependence of the modulation, some caution is in order
when fitting global spectral models.  We decided to divide the lightcurve
into 5 parts (shown in Fig.\ref{fig:lcfit}) labelled A (the eclipse)
through E. The detailed spectral fitting we have performed was carried out
for each of these phases to test the dependence of different parameters on
orbital phase. We have also investigated the dependence of the \pca\
colours on phase.  The strongest variations are found in the soft energy
bands [2.5--3.7 keV/3.7--5 keV and 3.7--5 keV/5--6.2 keV], where we find
variations of order 30\% to 50\%, with phase A being softest and phase D
being hardest.  In the more energetic bands [5--6.2 keV/6.2--10.2 keV and
6.2--10.2 keV/10.2--14.6 keV], the variation of colour with phase is
weaker, i.e., of order 10\%.  These results are consistent with the
analysis of \beppo\ data by P00; however, the signal is not nearly as
strong and as indicative as, for example, in the case of Her X-1. The
strong phase-dependent behaviour in that case has been interpreted as being
due to neutral hydrogen absorption and electron scattering in a large
column depth medium \cite{stelzer:98a}.

\begin{figure}
  \psfig{figure=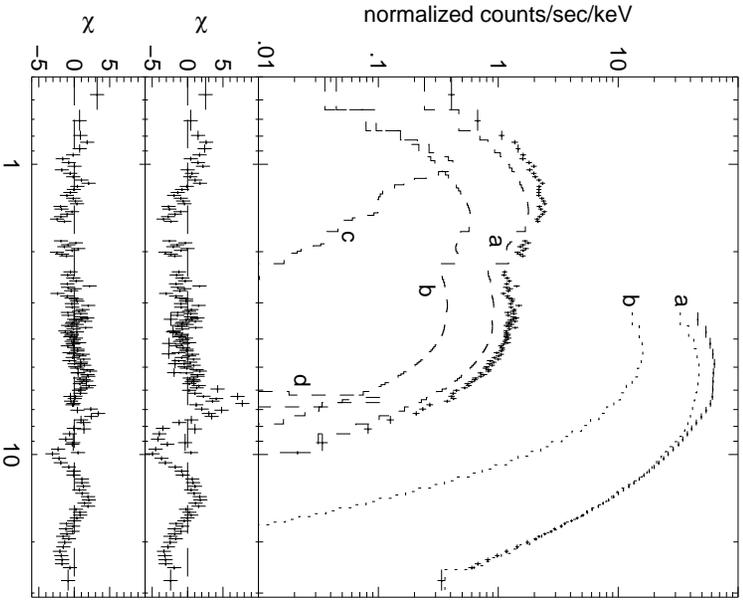,width=\columnwidth}
  \caption{Phase C spectrum and several best fit
    models. Top panel: spectrum and model components for the optically thin
    fit. Dotted line: PCA components; dashed line: SIS components. (a)
    cutoff powerlaw; (b) blackbody; (c) {\sf meka} plasma model; (d) gss.
    Second panel: blackbody plus cutoff power law; third panel: blackbody
    plus cutoff power law plus Gaussian plus Raymond-Smith plasma.}
  \label{fig:allmodels_thin}
\end{figure}

As a starting point we fitted test models to the complete data set (not
divided into phases). As with the previous analyses cited in the
introduction, simple one-component models cannot produce a satisfactory
fit--- neither a single blackbody, nor a thermal plasma, nor a cutoff power
law adequately fit the data.  Adding a second component can improve the fit
significantly. We found that the combination of a disc blackbody
\cite{mitsuda:84a} with either a simple blackbody or a cutoff power law
(plus a Gaussian to model the iron K${\alpha}$ line) did produce marginally
acceptable fits (residuals are plotted in Fig.\,\ref{fig:allmodels_thin}).
However, better fits can be achieved with physically more plausible models,
which are consistent with the strong constraints that light curve fitting
puts on the geometry of the system, which we will outline below.

If the central object in \eight\ is in fact a neutron star, any thermal
radiation from its surface is completely obscured by the accretion disc
(Hellier \& Mason 1989).  The observed X-rays are produced either in the
disc itself or in the corona.  Lightcurve fitting indicates that the
emission comes from a large extended source, and the (essentially required)
power law component in the spectra supports the notion that the corona is
the main radaition source.  We based our spectral fitting on this
assumption and investigated two possible, though not exclusive, spectral
models, one with an optically thick corona and one with an optically thin
corona.

\noindent{\em The optically thick case:} We expect the spectral signature
of the hot corona to be of power law shape, with an exponential high energy
cutoff produced by Comptonization, as is appropriate for many LMXBs
\cite{white:88a}.  (Here a cutoff at low energies would indicate a
relatively cool corona.) If the corona is optically thick, the spectrum
will not contain any component originating from the central object, as such
radiation will have been completely reprocessed.  The accretion disk is
nearly edge on and its atmosphere will partially cover the coronal
emission.  To account for this effect, we used a model consisting of a
partially absorbed cutoff power law and an iron line.  The partial
absorption is modelled by co-adding an unabsorbed and an absorbed cutoff
powerlaw with equal parameters.  The normalization of the absorbed
component, however, is multiplied by a constant, ${\rm const}_{\rm cpl}$,
which is related to the absorption fraction, $f_{\rm abs}$, by
\begin{equation}
        f_{\rm abs} \equiv \frac{{\rm const}_{\rm cpl}}{1 + {\rm
        const}_{\rm cpl}}~~. 
\end{equation} 
The absorption column in the partial absorption model is treated as a free
parameter, and we include both electron scattering and neutral hydrogen
absorption modelled using the cross sections of Ba\l{}uci\'{n}ska-Church \&
McCammon \shortcite{balu:92a}. Our spectral model is, of course, a
simplification because in any natural situation a range of absorbing
columns will occur.  Higher levels of detail, however, are not warranted
given our understanding of this source and the limits of the X-ray data.

Having found a satisfactory overall spectral fit, we used the phase
selected spectra to optimize the fits. We assumed that the physical
parameters in the emitting region do not change from phase to phase and
that the changes in the spectrum are entirely due to changes in obscuration
of the corona both by the completely opaque disk and the partial
absorption.  We thus tied the powerlaw slope and the cutoff energy to be
the same for all five phases, so that in the fitting process these
parameters vary in unison.  Similarly, we tied the iron line energy of all
five phases together and, for lack of high spectral resolution from the
\pca\ which dominates the statistics in this region, fixed the iron line
width to 0.1\,keV.  The foreground absorption column was also tied together
for all five phases.  Having combined the two \sis\ data sets and the two
\gis\ data sets, we were left with a total of 15 data sets, which we then
fit simultaneously.  The results of the joint fitting are presented in
Table\,\ref{tab:thick}.  The reduced $\chi^{2}$ is $\chi^{2}_{\rm red} =
0.78$.

It is worth noting that there is a correlation between the power law
normalization (overall count rate), the internal absorption column, and the
absorption fraction, with the internal absorption column being greatest in
the brightest phases.  Note also that the equivalent width of the iron line
is very large ($\sim 250$ eV, much larger than in the optically thin case).
A more detailed interpretation of these results will follow in
\S\ref{sec:discuss}.

\begin{figure}
  \psfig{figure=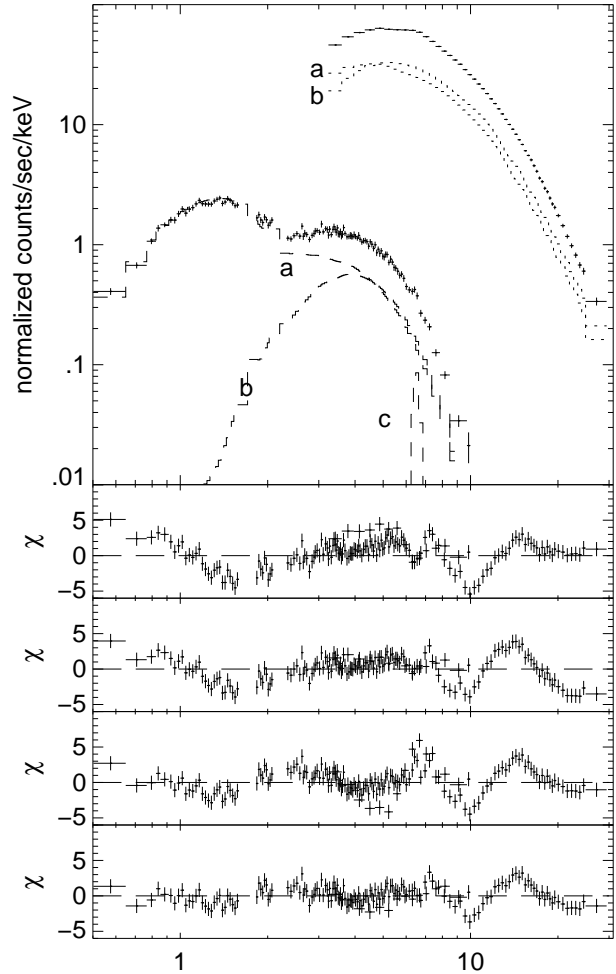,width=\columnwidth}
  \caption{Phase C spectrum and best fit models in the {\em optically
      thick} case. Top panel: spectrum and model components. Dotted line:
    {\em PCA} components; dashed line: {\em SIS} components. (a) Unabsorbed
    powerlaw; (b) absorbed powerlaw; (c) gaussian.  From second panel down:
    residuals of different models. In sequential order downward: (1) disc
    blackbody plus cutoff power law plus narrow Gaussian at $\sim 6.4$ keV;
    (2) disc blackbody plus blackbody plus narrow Gaussian at $\sim 6.4$
    keV; (3) partially absorbed cutoff powerlaw; (4) partially absorbed
    cutoff powerlaw plus iron line.  The parameters for the fit in the
    bottom panel are shown in Table \ref{tab:opt_thick}.}
  \label{fig:allmodels_thick}
\end{figure}

\begin{table}
\begin{center}
{\small \begin{sideways} \begin{minipage}{0.95\textheight}
{\small Model: ${\sf const.} \cdot {\sf phabs}(\rm NH_{\rm phabs})
\left [\left(1 + {\sf const._{cpl}}\cdot {\sf phabs}(NH_{\rm abs})
\cdot {\sf cabs}(NH abs)\right)
{\sf cutoffpl}(\Gamma_{\rm cpl},Ecut_{\rm cpl},N_{\rm cpl})
+ {\sf gauss}(E_{\rm gss},\sigma_{\rm gss},N_{\rm gss})\right]$} \\
\begin{tabular}{ccccccccccccccc}
\hline
\noalign{\vspace*{0.7mm}}
Phase &
$NH_{\rm phabs}$ &
$\Gamma_{\rm cpl}$ &
$Ecut_{\rm cpl}$ &
$N_{\rm cpl}$ &
const$_{\rm cpl}$ &
$NH_{\rm abs}$ &
$E_{\rm gss}$ &
$N_{\rm gss}$ &
$EW_{\rm gss}$ &
const$_{\rm SIS}$ &
const$_{\rm GIS}$ &
$\chi^2$ &
$\chi^2_{\rm red}$ \\
&
$(\times 10^{20})$&
&
(keV)&
$(\times 10^{-2})$&
	&
$(\times 10^{22})$&
(keV)&
$(\times 10^{-3})$&
(eV)&
&
&
&
&\\
\noalign{\vspace*{0.7mm}}
\hline
\noalign{\vspace*{0.7mm}}
A
& 
$\errtwa{7.95}{0.72}{0.71}$
& 
$\errtwa{0.80}{0.02}{0.02}$
& 
$\errtwa{11.76}{0.23}{0.22}$
& 
$\errtwa{1.82}{0.07}{0.07}$
& 
$\errtwa{0.89}{0.08}{0.08}$
& 
$\errtwa{2.51}{0.31}{0.29}$
& 
$\errtwa{6.51}{0.02}{0.02}$
& 
$\errtwa{0.66}{0.09}{0.08}$
& 
283
& 
$\errtwa{0.78}{0.02}{0.02}$
& 
$\errtwa{0.80}{0.02}{0.02}$
& 
& 
& 
\\ \noalign{\vspace*{2.7mm}}
B
& 
" & 
" & 
" & 
$\errtwa{3.44}{0.09}{0.08}$
& 
$\errtwa{1.15}{0.06}{0.06}$
& 
$\errtwa{4.97}{0.32}{0.31}$
& 
" & 
$\errtwa{1.44}{0.18}{0.17}$
& 
328
& 
$\errtwa{0.75}{0.01}{0.01}$
& 
$\errtwa{0.79}{0.01}{0.01}$
& 
& 
& 
\\ \noalign{\vspace*{2.7mm}}
C
& 
" & 
" & 
" & 
$\errtwa{3.81}{0.09}{0.09}$
& 
$\errtwa{1.36}{0.06}{0.06}$
& 
$\errtwa{5.19}{0.28}{0.28}$
& 
" & 
$\errtwa{1.31}{0.15}{0.14}$
& 
270
& 
$\errtwa{0.74}{0.01}{0.01}$
& 
$\errtwa{0.80}{0.01}{0.01}$
& 
3636
& 
0.781
\\ \noalign{\vspace*{2.7mm}}
D
& 
" & 
" & 
" & 
$\errtwa{3.15}{0.07}{0.07}$
& 
$\errtwa{1.21}{0.06}{0.06}$
& 
$\errtwa{3.77}{0.19}{0.18}$
& 
" & 
$\errtwa{1.05}{0.05}{0.05}$
& 
261
& 
$\errtwa{0.77}{0.01}{0.01}$
& 
$\errtwa{0.76}{0.01}{0.01}$
& 
& 
& 
\\ \noalign{\vspace*{2.7mm}}
E
& 
" & 
" & 
" & 
$\errtwa{2.32}{0.06}{0.06}$
& 
$\errtwa{1.16}{0.06}{0.06}$
& 
$\errtwa{3.56}{0.26}{0.25}$
& 
" & 
$\errtwa{0.70}{0.10}{0.10}$
& 
239
& 
$\errtwa{0.74}{0.01}{0.01}$
& 
$\errtwa{0.79}{0.01}{0.01}$
& 
& 
& 
\\ \noalign{\vspace*{2.7mm}}
\end{tabular}

\caption{\label{tab:opt_thick}Best fit parameters for the `optically thick
  model'. Again, the physical parameters are tied together (power law slope
  and cutoff, iron line energy), denoted by the symbol ". The reduced
  $\chi^2$ of this fit is ${\chi^{2}}_{\rm red} = 0.78$, with 4656
  degrees of freedom.  The iron line width was frozen at 0.1\,keV. 
  \protect{\label{tab:thick}}}
\end{minipage} \end{sideways}}
\end{center}
\end{table}

\noindent{\em The optically thin case:} What distinguishes the optically
thin from the optically thick case is that radiation from the central
source can be scattered into the observer's line of sight by the corona,
without having its input spectrum being completely distorted.  One might
therefore expect to see a blackbody component in combination with a cutoff
power law.  In principle, both of these components will be partially
absorbed, as assumed in the optically thick case.  Since the opically thick
model already fits the data satisfactorly, however, such a model would be
an over-parametrization given the limits of the data.  To highlight the
differences between the two models, we therefore neglected partial
absorption in our fits that included blackbody emission.

Such a model can fit the data reasonably well, leaving residuals at both
the low and high energy ends. The low energy residuals could be caused by a
line complex, as also would be expected to emanate from an optically thin
(probably photoionized) corona. We have modelled these low energy residuals
with the {\tt meka} plasma model; (Kaastra \nocite{kaastra:92a}
1992, and references therein).  Once again, we fitted the five phase
selected spectra simultaneously, and we constrained the temperatures of the
blackbody and the thermal plasma to be uniform throughout all five phases.
We also constrained the line energy and both the power law index and cutoff
energy to be uniform throughout the phases.  The resulting best fit with
${\chi^2}_{\rm red} = 0.74$ is presented in Table\,\ref{tab:thin} and in
Fig.\,\ref{fig:allmodels_thin}.

We note that the lack of absorption in the best fit model is somewhat
inconsistent with the predicted {\tt COLDEN} NH value.  Forcing the
hydrogen column to the {\tt COLDEN} value worsens the fit slightly, leading
to residuals below 0.8 keV.  Since the exact distance to \eight\ is unknown
and the lack of partial absorption in the optically thin case is an
idealization anyway, we decided to leave the clarification of this issue to
{\em Chandra} and {\em XMM} observations.

\begin{table}
\begin{center}
{\small \begin{sideways} \begin{minipage}{0.95\textheight}
{\small Model: ${\sf const.} \cdot {\sf phabs}(\rm NH_{\rm phabs})
\left[{\sf bbody}(T_{\rm bbd},N_{\rm bbd}) + 
{\sf cutoffpl}(\Gamma_{\rm cpl},Ecut_{\rm cpl},N_{\rm cpl})
+ {\sf gauss}(E_{\rm gss},\sigma_{\rm gss},N_{\rm gss})
+ {\sf raymond}(T_{\rm ray},1,0,N_{\rm ray})\right]$} \ \ \ \
$\chi^2/{\rm dof} = \frac{3452}{4665} = 0.74$ \\
\begin{tabular}{cccccccccccccccc}
\hline
\noalign{\vspace*{0.7mm}}
Phase &
$NH_{\rm phabs}$ &
$kT_{\rm bbd}$ &
$N_{\rm bbd}$ &
$\Gamma_{\rm cpl}$ &
$Ecut_{\rm cpl}$ &
$N_{\rm cpl}$ &
$E_{\rm gss}$ &
$N_{\rm gss}$ &
$EW_{\rm gss}$ &
$kT_{\rm ray}$ &
$N_{\rm ray}$ &
const$_{\rm SIS}$ &
const$_{\rm GIS}$ \\
&
$(\times 10^{20})$&
(keV)&
$(\times 10^{-3})$&
&
(keV)&
$(\times 10^{-3})$&
(keV)&
$(\times 10^{-4})$&
(eV)&
(keV)&
$\times 10^{-3}$
&
&
\\
\noalign{\vspace*{0.7mm}}
\hline
\noalign{\vspace*{0.7mm}}
A
& 
$\errtwa{0.00}{0.10}{0.00}$
& 
$\errtwa{1.39}{0.01}{0.01}$
& 
$\errtwa{1.44}{0.03}{0.03}$
& 
$\errtwa{0.04}{0.00}{0.50}$
& 
$\errtwa{7.95}{0.03}{0.03}$
& 
$\errtwa{0.77}{0.01}{0.01}$
& 
$\errtwa{6.48}{0.02}{0.02}$
& 
$\errtwa{0.44}{0.08}{0.08}$
& 
101
& 
$\errtwa{1.01}{0.03}{0.04}$
& 
$\errtwa{3.23}{0.52}{0.42}$
& 
$\errtwa{0.79}{0.01}{0.01}$
& 
$\errtwa{0.81}{0.02}{0.02}$
\\ \noalign{\vspace*{2.7mm}}
B
& 
" & 
" & 
$\errtwa{2.04}{0.05}{0.04}$
& 
" & 
" & 
$\errtwa{1.70}{0.01}{0.01}$
& 
" & 
$\errtwa{1.40}{0.13}{0.12}$
& 
160
& 
" & 
$\errtwa{4.41}{0.63}{0.48}$
& 
$\errtwa{0.75}{0.01}{0.01}$
& 
$\errtwa{0.78}{0.01}{0.01}$
\\ \noalign{\vspace*{2.7mm}}
C
& 
" & 
" & 
$\errtwa{2.20}{0.06}{0.05}$
& 
" & 
" & 
$\errtwa{2.07}{0.01}{0.01}$
& 
" & 
$\errtwa{1.73}{0.14}{0.14}$
& 
167
& 
" & 
$\errtwa{3.21}{0.59}{0.43}$
& 
$\errtwa{0.74}{0.00}{0.00}$
& 
$\errtwa{0.78}{0.01}{0.01}$
\\ \noalign{\vspace*{2.7mm}}
D
& 
" & 
" & 
$\errtwa{2.28}{0.02}{0.03}$
& 
" & 
" & 
$\errtwa{1.58}{0.01}{0.01}$
& 
" & 
$\errtwa{1.02}{0.05}{0.04}$
& 
121
& 
" & 
$\errtwa{3.50}{0.58}{0.34}$
& 
$\errtwa{0.76}{0.01}{0.01}$
& 
$\errtwa{0.75}{0.01}{0.01}$
\\ \noalign{\vspace*{2.7mm}}
E
& 
" & 
" & 
$\errtwa{1.62}{0.04}{0.03}$
& 
" & 
" & 
$\errtwa{1.15}{0.01}{0.01}$
& 
" & 
$\errtwa{0.68}{0.10}{0.09}$
& 
111
& 
" & 
$\errtwa{2.34}{0.46}{0.30}$
& 
$\errtwa{0.74}{0.01}{0.01}$
& 
$\errtwa{0.79}{0.01}{0.01}$
\\ \noalign{\vspace*{2.7mm}}

\end{tabular}

\caption{\label{tab:opt_thin}Best fit parameters for the `optically thin
  model'. Physical parameters (blackbody temperature, power law slope and
  cutoff, plasma temperature, and iron line energy) are tied to stay
  constant as a function of phase, denoted by the symbol ". The reduced
  $\chi^{2}$ for this fit is ${\chi^{2}}_{\rm red} = 0.74$, with 4655
  degrees of freedom.  We have frozen the iron line width to $0.1$\,keV.
  Similarly, redshift and metallicity of the Raymond-Smith component were
  frozen to their canonical values (0 and 1 respectively).
  \protect{\label{tab:thin}}}
\end{minipage} \end{sideways}}
\end{center}
\end{table}

\begin{figure}
  \psfig{figure=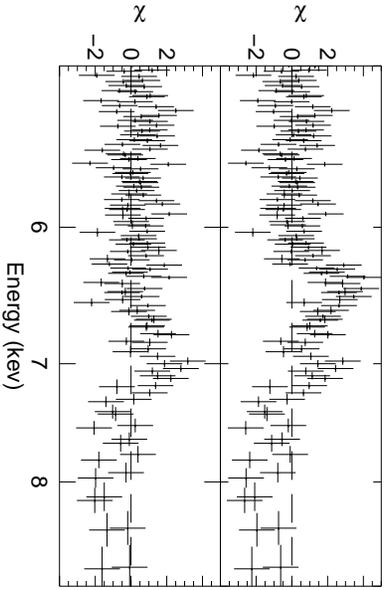,width=\columnwidth}
  \caption{Residuals for a model with only a cutoff power law to model the
    continuum (top panel), and for the same model plus a narrow Gaussian
    iron line at 6.4 keV (bottom panel).  \label{fig:iron_line}}
\end{figure}  

In both cases, the high energy end of the spectrum shows some systematic
residuals, present also in the fits based on disc emission models. We
suspect that at least some of this effect can be attributed to systematic
errors in the \rxte\ response, in particular, differences in the high
energy slopes of \asca\ and \rxte. However, the \ginga\ data reported by
Hellier et al. \shortcite{hellier:92a} also seem to show this hard excess,
which would argue against a purely instrumental effect. In this paper we
will not attempt a physical explanation of this feature. The failure of
\astroe\ with its high energy capabilities is particularly unfortunate
in this respect.

We then tried to determine the structure of the iron line.  Since the
sensitivity of the \sis\ has been deteriorating, we decided to use only the
1993 \sis\ dataset from the 1993 \asca\ observation. We limited the fits to
the energy range from 3--10\,keV and approximated the continuum spectrum
around the line by a cutoff power law. Two narrow lines produce the most
significant improvement in $\chi^2$, from 587/458 to 421/454, in agreement
with P00.  The line energies are 6.4\,keV and 7\,keV, consistent with a
cold K${\alpha}$ line and either K${\beta}$ or ionized K${\alpha}$.  The
equivalent widths of the two lines are on the order of 80\,eV and 40\,eV
respectively.

\section{Modelling the Lightcurve}\label{sec:lcfit}

Models of the \eight\ lightcurve, for both the optical and X-ray energy
bands, have been presented by White \& Holt \shortcite{white:82a}, Mason \&
C\'ordova \shortcite{mason:82a}, and Hellier \& Mason
\shortcite{hellier:89a}. Essentially, these models have taken emission from
an extended central region (X-rays) plus emission from a disc and disc rim
(optical), and then considered the effects of obscuration by both the
secondary star (assumed to be Roche lobe filling) and by the accretion disc
rim.  The disc rim has been modelled by placing nodes symmetrically above
and below the disc mid-plane at a fixed disc radius but at variable heights
and phases along the disc edge. The disc rim profile is then linearly
extrapolated at intermediate phases. Fit parameters for these models have
included the radius of the X-ray emitting corona, usually assumed spherical
for simplicity, the heights and phases of the rim nodes, the mass of the
compact object, and the inclination angle of the system with respect to our
line of sight (see Hellier \& Mason \nocite{hellier:89a} 1989).

Here we elaborate upon these models in two ways.  First, due to the
excellent statistics of \rxte, we are able to divide the \rxte\ lightcurves
into five distinct energy bands covering \pca\ pha channels 8--11, 12--15,
16--19, 20--27, and 28--32 (2.9--4.4\,keV, 4.4--5.8\,keV, 5.8--7.2\,keV,
7.2--10.2\,keV, and 10.2--12.0\,keV, respectively). We fit each of these
energy band lightcurves simultaneously. Second, we model both an optically
thin and an optically thick spherical corona.  Specifically, we model the
X-ray emission via a uniform emissivity or a uniform surface brightness.
We do not expect (and as we confirm below; see also Hellier \& Mason
\nocite{hellier:89a} 1989) strong differences between the two cases.  Given a
completely edge-on viewing angle, for a given disc rim height the fraction
of emission that is obscured and the emission-weighted mean height of 
the unobscured radiation are nearly identical for both uniform surface
brightness and uniform emissivity.  In our fits, we allow both the overall
flux normalization and the radius of the sphere to vary for each energy
band; however, the disc rim parameters (radius of the disc, $R_{\rm rim}$,
node heights, $H$, and node phases, $\phi$) are held fixed.  Additional fit
parameters are the mass of the compact object ($M_1$) and the inclination
of the system with respect to our line of sight ($i$).

In part these parameters are determined by utilizing the fact that \eight\ 
has a known binary period of $P=5.57$\,hr and a measured projected primary
orbital velocity of $K_1=70~{\rm km s^{-1}}$ \cite{mason:82b,cowley:82a}.
We highlight the systematic uncertainties in our fit parameters by varying
this latter parameter by $+ 30~{\rm km s^{-1}}$ (see Cowley et al.
\nocite{cowley:82a} 1982). Furthermore, we search for model fits with
$M_1 \sim 0.3$, $1.4$ and $2.5~\msun$. That is, we fit models for a `white
dwarf', `canonical neutron star', and `low mass black hole' primary.  The
latter primary mass yields a secondary mass consistent with Roche lobe
overflow from a main sequence star (see below).

In order to better interpret the results of our fits, we first present
estimates of the characteristic accretion system size and mass scales.
Defining the secondary to primary mass ratio, $q \equiv M_2/M_1$, Newton's
laws give $q \approx 0.21$ for $M_1 = 1.4~\msun$, $K_1 = 70~{\rm km
  s^{-1}}$, and $i=85^\circ$.  Henceforth, these parameter values shall be
referred to as the `nominal parameters'.  As noted by Mason et al.
\shortcite{mason:82b}, this mass ratio is roughly half that expected from
the mass-period relationship for a lower main sequence secondary. If
instead we choose $M_1 = 2.5~\msun$, $K_1 = 100~{\rm km~s^{-1}}$, and
$i=85^\circ$, one obtains a secondary mass of $M_2 = 0.62~\msun$, as one
would expect for the period-mass relationship of a main sequence star
\cite{accretion}.  The above system parameters yield binary separations of
$a \approx 1.3$--$1.6 \times 10^{11}$\,cm, respectively.  Over
the parameter ranges of interest to us, $q$ varies roughly as $(K_1/\sin i)
M_1^{-1/3}$, and the binary separation weakly varies as
$(1+q)^{1/3}~M_1^{1/3}$.

The disc circularization radius can be determined via the approximation
\begin{equation}
R_{\rm circ} = (1+q) (0.5 - 0.227 \log q )^4 ~ a
\label{eq:rcirc}
\end{equation}
\cite{accretion}, where for the nominal parameters $R_{\rm circ} \approx
0.22a \approx 2.9 \times 10^{10}$\,cm. As a fraction of the binary
separation, this ratio is fairly constant for the parameters of concern to
us.  The disc tidal truncation radius is approximately given by $R_{\rm T}
\approx 0.9~R_1$, where $R_1$ is the Roche lobe radius of the primary.
Approximating this radius \cite{eggleton:83a} as
\begin{equation}
\frac{R_1}{a} \approx \frac { 0.49 q^{-2/3} }{0.6 q^{-2/3} +
  \ln(1+q^{-1/3})}
\label{eq:roche} 
\end{equation}
yields $R_{\rm T} \approx 0.47~a \approx 6.1 \times 10^{10}$\,cm for the
nominal parameters. Again, expressed as a fraction of the binary
separation, this radius is relatively constant for the parameters of
interest to us.

Equation~\ref{eq:roche}, with $q$ replaced by $q^{-1}$, also yields $R_2/a$
($\sim 0.22$ for the nominal parameters), where $R_2$ is the Roche lobe
radius of the secondary.  For a near edge on system with a brief (relative
to the orbital period) eclipse, $R_2/a$ is approximately proportional to
the fraction of the binary period that can be eclipsed by the secondary.
This fraction is roughly proportional to $q^{1/3} \propto M_1^{-1/9}
(K_1/\sin i)^{1/3}$. As the duration of the eclipse is a fixed fraction of the
binary orbital period, we therefore expect the coronal radius, expressed as
a fraction of the binary separation, to be weakly dependent upon $K_1/\sin
i$ and almost completely independent of $M_1$ (see also Mason \& C\'ordova
1982).  As we show below, we indeed can fit a relatively wide range for
$M_1$\footnote{If we insist that the (main sequence) radius of the
  secondary not be greater than $R_2$, then we require $M_1 \aproxlt 2.5$.
  Hence we only consider primary masses of $\approx 0.3$, $1.4$, and
  $2.5~\msun$.}. The mass ratio, $q$, however, is more tightly constrained
by the lightcurve fits.

We divide the lightcurves into 50 phase bins in each energy band.  In lieu
of statistical errors, which were very small given the large effective area
of \rxte, a systematic error of 4.2\% was added to each phase bin.  This
represented the average variance of the lightcurve from orbit to orbit over
the \eight\ period. (Approximately four orbital periods were measured in
part or in whole.)  For the optically thin case, the corona was divided
into 6 evenly spaced radial zones, 50 evenly spaced zones in the azimuthal
angle, $\phi$, and 50 zones evenly spaced in $\mu \equiv \cos \theta$ in
the polar angular direction, $\theta$. The optically thick corona had
$50\times50$ evenly spaced zones in $\phi$ and $\mu$. We assumed a disc rim
with $9\times2$ nodes symmetrically placed above and below the disc
mid-plane.  Ray tracing was performed to determine whether emission from a
given coronal element intercepted the disc rim.  Near zero phase, the {\tt
  blink} subroutine from Keith Horne's eclipse mapping code
\cite{horne:85a} was used to determine if the secondary was blocking the
line of sight to the emission element.  In all, there were 31 fit
parameters (primary mass, system inclination, 5 coronal radii and
emissivities/surface brightnesses, disc rim radius, and 9 node heights and
phases). For the fits, all energy bands were weighted equally, and the {\tt
  amoeba} subroutine from Press et al.  \shortcite{nr} was used in the
$\chi^2$ minimization. Fit results are presented in
Tables~\ref{tab:lcfit_a} and \ref{tab:lcfit_b}.

\begin{table*}
\caption{\small System and Coronal Parameters for Fits to the \eight\
  \rxte\ Lightcurves. All length scales are in units of the binary
  separation, $a$. $R_{c_{a\_b}}$ is the radius of the optically thin or
  thick, uniformly emitting corona for \rxte\ pha channels
  $a$--$b$. $R_{\rm rim}$ is the radius of the obscuring rim.  $M_1$
  is the mass of the central compact object, and $i$ is the angle of
  the normal to the binary orbital plane with respect to our line of
  sight. Values given are assuming a projected primary velocity of $K_1$.
  \protect{\label{tab:lcfit_a}}}
\center{
\begin{tabular}{lcccccccccc}
\hline
\noalign{\vskip 3pt}
Model & $M_1$ & $K_1$ & $i$ & $R_{c_{8\_11}}$ & $R_{c_{12\_15}}$ &
$R_{c_{16\_19}}$ & $R_{c_{20\_27}}$ & $R_{c_{28\_32}}$ & $R_{\rm rim}$
& $\chi^2$/dof \\
& (${\rm M}_{\odot}$) & (${\rm km~s^{-1}}$) & ($^\circ$) \\
\noalign{\vskip 3pt}
\hline
\noalign{\vskip 3pt}
Thin &  {0.32} & {70} & {80.7} &  {0.220} &  {0.216} &
 {0.210} &  {0.211} &  {0.211} &  {0.38} & 188/219 \\
\noalign{\vskip 3pt}
Thin &  {1.40} & {70} & {83.4} &  {0.219} &  {0.217} &
 {0.216} &  {0.216} &  {0.215} &  {0.41} & 154/219 \\
\noalign{\vskip 3pt}
Thin &  {2.47} & {100} & {82.5} &  {0.220} &  {0.216} &
 {0.213} &  {0.213} &  {0.212} &  {0.40} & 175/219 \\
\noalign{\vskip 3pt}
\noalign{\vskip 12pt}
Thick & {0.31} & {70} &  {81.1} &  {0.216} &  {0.213} &
 {0.209} &  {0.210} &  {0.210} &  {0.40} & 181/219 \\
\noalign{\vskip 3pt}
Thick & {1.38} & {70} &  {83.7} &  {0.215} &  {0.214} &
 {0.213} &  {0.213} &  {0.213} &  {0.39} & 180/219 \\
\noalign{\vskip 3pt}
Thick & {2.48} & {100} &  {83.0} &  {0.216} &  {0.213} &
 {0.212} &  {0.212} &  {0.212} &  {0.41} & 180/219 \\
\noalign{\vskip 3pt}
\end{tabular}
}
\end{table*}

\begin{table*}
\caption{\small Disc Rim Parameters for Fits to the \eight\
  \rxte\ Lightcurves. All length scales are in units of the binary
  separation, $a$, and phases are in units of the binary phase. $\phi$ and $H$
  are, respectively, the phase and height of disc rim nodes $N_1$--$N_9$.
  Rim heights inbetween the nodes are determined by linear interpolation.
  Values given are assuming a projected primary velocity of
  $K_1$. \protect{\label{tab:lcfit_b}}} 
\center{
\begin{tabular}{lccrccccccccc}
\noalign{\vskip 3pt}
Model & $M_1$ & $K_1$ & & $N_1$ & $N_2$ & $N_3$ & $N_4$ & $N_5$ & $N_6$ &
$N_7$ & $N_8$ & $N_9$ \\ 
\noalign{\vskip 3pt}
& ($\msun$) & (${\rm km~s^{-1}}$) \\
\noalign{\vskip 3pt}
\hline
\noalign{\vskip 3pt}
Thin & {0.32} & {70} & $\phi=$ & 0.111 & 0.295 & 0.321 & 0.408 & 0.526 &
0.750 & 0.813 & 0.934 & 0.943 \\ 
& & & $H=$ & 0.110 & 0.093 & 0.075 & 0.102 & 0.099 & 0.136 & 0.132 & 0.109 &
0.108 \\
\noalign{\vskip 3pt}
Thin & {1.40} & {70} & $\phi=$ & 0.107 & 0.202 & 0.317 & 0.399 & 0.564 &
0.741 & 0.846 & 0.882 & 0.938 \\ 
& & & $H=$ & 0.104 & 0.105 & 0.080 & 0.097 & 0.103 & 0.129 & 0.132 & 0.096 &
0.117 \\
\noalign{\vskip 3pt}
Thin & {2.47} & {100} & $\phi=$ & 0.124 & 0.299 & 0.317 & 0.413 & 0.537 &
0.768 & 0.830 & 0.889 & 0.907 \\ 
& & & $H=$ & 0.106 & 0.090 & 0.078 & 0.099 & 0.100 & 0.133 & 0.132 & 0.100 &
0.112 \\
\noalign{\vskip 3pt}
\noalign{\vskip 12pt}
Thick & {0.31} & {70} & $\phi=$ & 0.112 & 0.291 & 0.323 & 0.410 & 0.535 &
0.750 & 0.831 & 0.924 & 0.928 \\ 
& & & $H=$ & 0.111 & 0.090 & 0.074 & 0.101 & 0.100 & 0.139 & 0.134 & 0.103 &
0.108 \\
\noalign{\vskip 3pt}
Thick & {1.38} & {70} & $\phi=$ & 0.103 & 0.292 & 0.359 & 0.441 & 0.553 &
0.756 & 0.828 & 0.896 & 0.911 \\ 
& & & $H=$ & 0.111 & 0.088 & 0.082 & 0.103 & 0.101 & 0.137 & 0.132 & 0.104 &
0.112 \\
\noalign{\vskip 3pt}
Thick & {2.48} & {100} & $\phi=$ & 0.111 & 0.296 & 0.343 & 0.419 & 0.537 &
0.765 & 0.819 & 0.922 & 0.930 \\ 
& & & $H=$ & 0.109 & 0.088 & 0.079 & 0.100 & 0.101 & 0.137 & 0.132 & 0.104 &
0.110 \\
\noalign{\vskip 3pt}
\end{tabular}
}
\end{table*}

\begin{figure}
\centerline{
\includegraphics[width=0.47\textwidth]{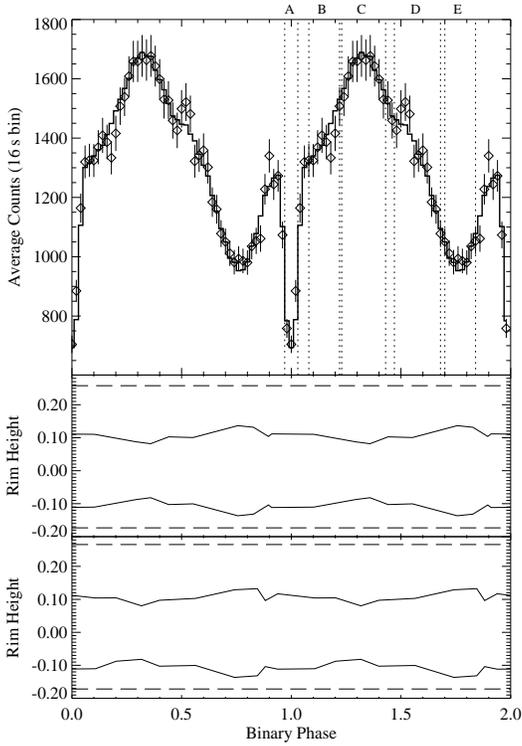}
}
\caption{\small   {\it Top panel:} Folded \eight\
  lightcurve in \rxte\ pha channels 12--15 (4.4--5.8\,keV) plus the best
  fit model (solid line) for a uniformly emitting, optically thick corona
  obscured by a disc rim. (Dotted lines and labels A--E indicate the
  regions of the phase-resolved spectral fits.)  {\it Middle panel:} The
  best fit disc rim profile, in units of the binary separation.  The dashed
  lines show the projected positions (relative to the disc rim, with
  respect to our line of sight) of the coronal poles.  {\it Bottom panel:}
  The same as the middle panel, but instead for the best fit optically thin
  corona model.  {\it Note:} Both of the above models assume a projected
  primary velocity of $K_1=70~{\rm km~s^{-1}}$ and $M_1 \approx 1.4~\msun$.
  Animations of the \eight\ system and lightcurve can be found at {\tt
    http://rocinante.colorado.edu/\~\,heinzs/1822/.}
  \protect{\label{fig:lcfit}}}
\end{figure}

Fig.~\ref{fig:lcfit} shows the best fit optically thick corona model for
the 4.4--5.8\,keV energy band, as well as the fitted disc rim profiles for
both the optically thick and thin models, both for $M_1 \sim 1.4~\msun$,
$K_1 = 70$\,${\rm km~s^{-1}}$.  (Animations of the fits can be viewed at
\hfill {\tt http://rocinante.colorado.edu/\~\,heinzs/1822/.}) We note that
using fewer coronal grid points did not adequately resolve the coronal
emission; however, numerical discreteness in the fitting process led to
uncertainties of $\Delta \chi^2 \approx 2$ in any fits with a greater
number of coronal grid points. In Table~\ref{tab:lcfit_a} we therefore only
present the systematic errors associated with varying the projected primary
velocity, $K_1$, between $70$ and $100~{\rm km~s^{-1}}$ and from searching
for $\chi^2$ minima near primary masses of $0.3$, $1.4$, and $2.5~\msun$.
We also experimented with varying the number of nodes in the fit; however,
9 nodes were sufficient to produce a reduced $\chi^2\aproxlt1$.

Both the optically thick and optically thin coronal models fit the data
nearly equally well, with $\chi^2$ ranging from $154$--$188$ for 219
degrees of freedom.  Although there is a slight preference for an optically
thin corona with $M_1 \approx 1.4~\msun$, given the simplicity of our
assumptions we do not consider these differences to be strongly
significant.

Most of the fits showed a weak trend for the coronal radius to decrease
with energy, as would be expected from the energy-dependent fractional
modulation shown in Fig.\,\ref{fig:depths}. The fractional change in the
coronal radii was as large as 5\%; however, this is still somewhat smaller
than the $\sim 10\%$ found in the two energy channel (1--6\,keV and
6--30\,keV) \ginga\ data fits of Hellier et al. \shortcite{hellier:92a}.
The other size scale parameters are consistent with previous models of the
lightcurve \cite{white:82a,mason:82a,hellier:89a}.  Specifically, the disc
rim height shows a minimum near phase 0.3--0.35, and a maximum near phase
0.75--0.85.  The disc rim height is $\approx 0.1a\approx
1.5\times10^{10}$\,cm.  The radius of the disc rim is consistent with being
at the tidal truncation radius, which is reasonable if the disc rim is due
to the interaction of the secondary's accretion stream with the disc (see,
for example, Armitage \& Livio \nocite{armitage:96a,armitage:98a} 1996,
1998, and references therein). The coronal radii are all $\approx
0.2a\approx 3\times10^{10}$\,cm, i.e.  consistent with the corona extending
all the way to the disc circularization radius.

We find a system inclination of $i \approx 81^\circ$--$84^\circ$, which is
also consistent with previous models of the lightcurve \cite{hellier:89a}.
As expected from the analysis discussed above, the fits for $M_1 \approx
0.3$, $1.4$, and $2.5~\msun$ were nearly identical.  Thus, based upon the
lightcurve analysis alone, it is impossible to determine whether \eight\ is
a white dwarf, neutron star, or a black hole.  Further observations,
especially any that can independently measure the velocity of the
secondary, are required to break the degeneracy in the primary mass values.
We note that to this end, Harlaftis et al.  \shortcite{harlaftis:97a} place
a lower limit on the companion velocity of $K_2 \aproxgt 225$\,km~s$^{-1}$.
If the lower end of this value is accurate and $K_1=70$\,km~s$^{-1}$, then
$q = K_1/K_2 \approx 0.3$ (consistent with the above estimates) and
$M_1+M_2 \approx 0.6~\msun \propto (K_1+K_2)^3$, i.e., \eight\ is a white
dwarf system.

\section{Variability Analysis}\label{sec:var}

Extremely high time resolution data were available from the \rxte\ 
observations.  Accordingly, we created lightcurves with $2^{-10}$\,s
resolution in order to search for high-frequency variability. No
significant variability was detected at frequencies $\aproxgt 1$\,Hz;
therefore, we created a series of lightcurves with 0.5\,s resolution.  We
chose three energy bands covering PCA pha channels 5--10, 16--19, and
28--41 (1.8--4\,keV, 5.8--7.2\,keV, and 10.2--15.3\,keV
respectively).  These energy bands were chosen to represent a low energy
band, an Fe K$\alpha$/$\beta$ band, and a high energy band.  
Energies higher than $\approx 15$\,keV were background dominated. 

\begin{figure*}
\centerline{
\includegraphics[width=0.33\textwidth]{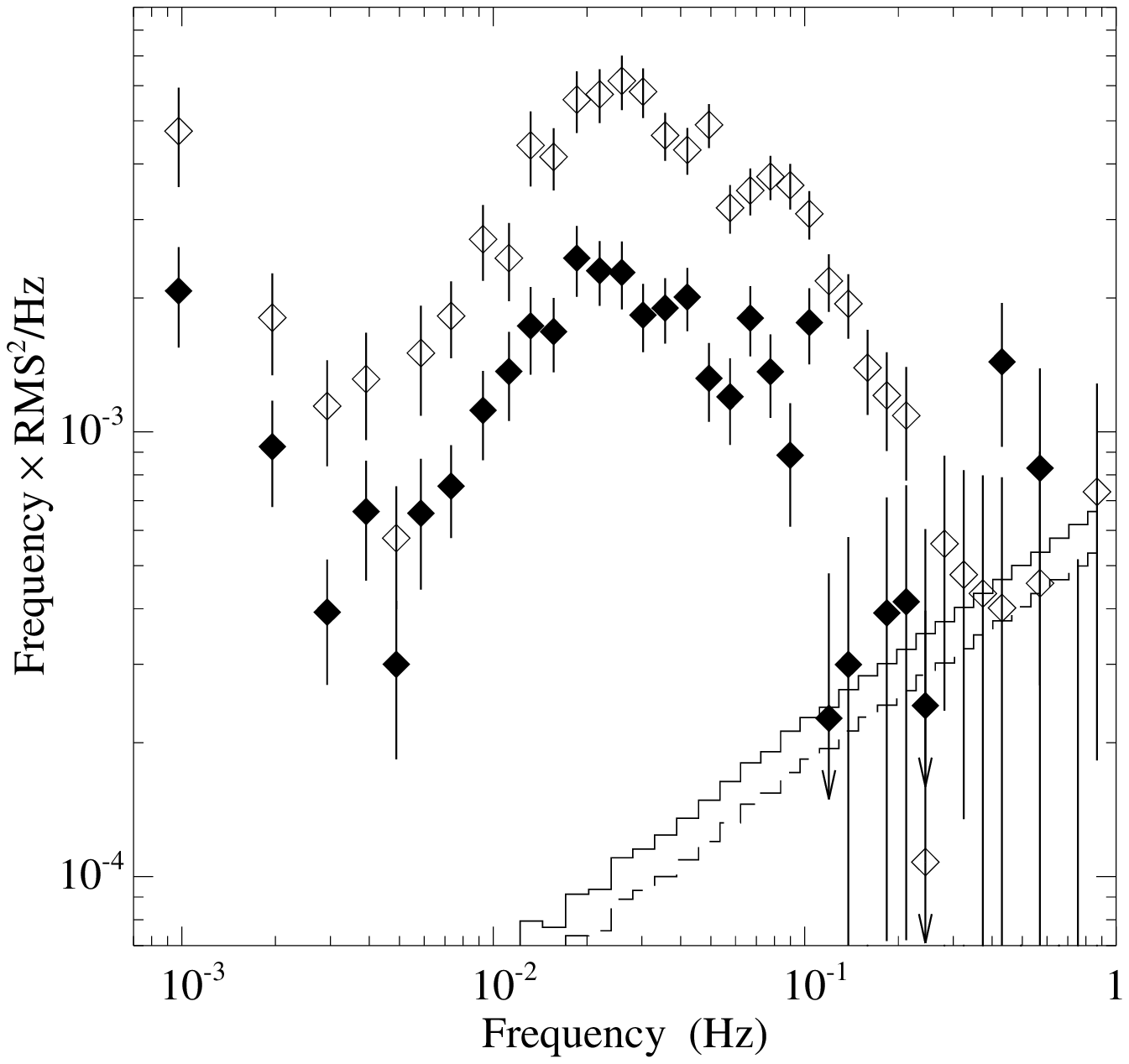}
\includegraphics[width=0.33\textwidth]{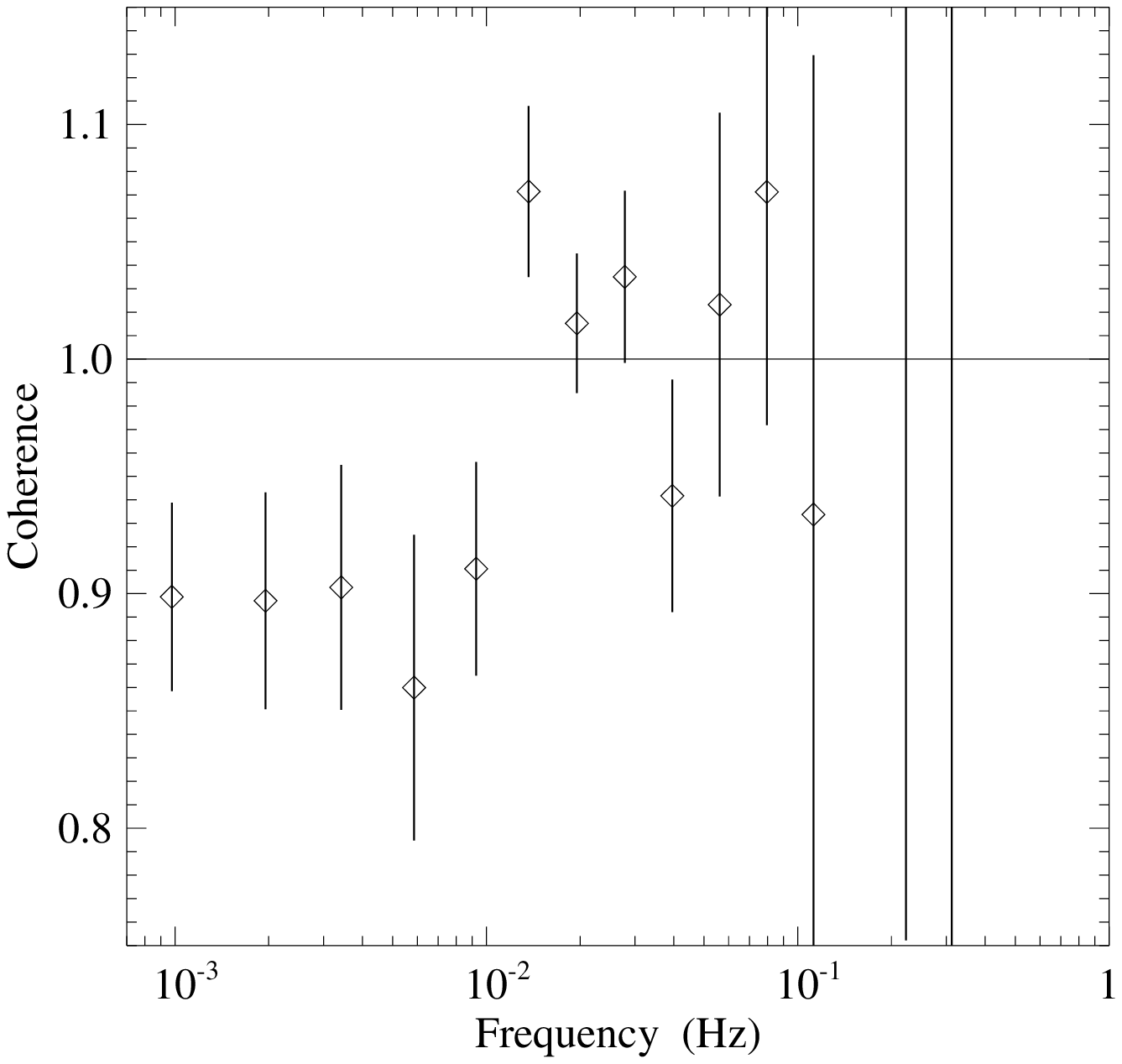}
\includegraphics[width=0.33\textwidth]{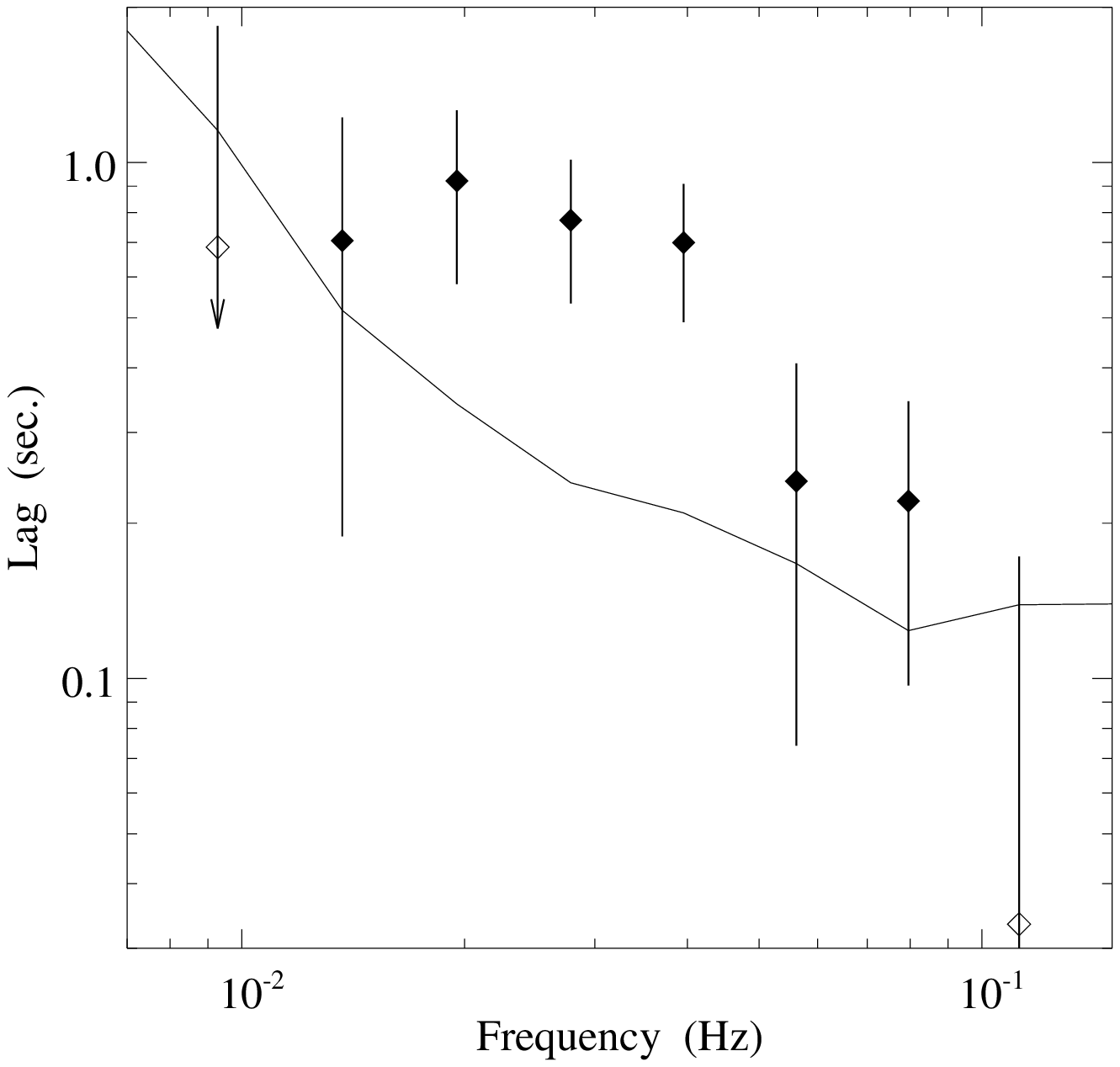}
}
\caption{\small {\it Left:} Power spectral density times Fourier frequency
  in the 1.8--4\,keV and 10.2--15.3\,keV energy bands (filled and clear
  diamonds, respectively). The solid line (1.8--4\,keV) and dashed line
  (10.2--15.3\,keV) show the residual noise levels, which correspond to the
  expected amplitudes of positive 1-$\sigma$ fluctuations above the mean
  value of the Poisson noise PSDs. {\it Centre:} Coherence function between
  the variability in the 1.8--4\,keV and 10.2--15.3\,keV energy bands. {\it
    Right:} Fourier frequency-dependent time lags between variability in
  the 1.8--4\,keV and 10.2--15.3\,keV energy bands. Filled diamonds
  indicate the soft band lagging the hard band, and clear diamonds indicate
  the hard band lagging the soft band.  The solid line corresponds to the
  magnitude of the expected 1-$\sigma$ Poisson noise level.
  \protect{\label{fig:psd}}}
\end{figure*}

We consider three measures of the variability: the power spectral density
(PSD) in each energy band, and the Fourier frequency-dependent time lags
and coherence function between variability in different energy bands.  A
discussion of Fourier techniques in specific, and timing analysis in
general, has been presented by {van der Klis} \shortcite{vanderklis:89b}.
Here we apply these Fourier analysis techniques in the same manner as for
our \rxte\ observations of Cyg X--1 \cite{nowak:99a}. Specifically, we used
the same techniques for estimating: deadtime corrections
\cite{zhangw:95a,zhangw:96a}; the error bars and Poisson noise levels of
the PSD \cite{leahy:83a,vanderklis:89b}; the error bars and noise levels
for the coherence function \cite{bendat,vaughan:97a}; and the error bars
and noise levels for the Fourier frequency-dependent time lag between hard
and soft photon variability \cite{bendat,nowak:99a}.  Results of these
analyses for the lowest and highest energy bands are presented in
Fig.~\ref{fig:psd}.  In this figure, the PSDs are normalized according to
Belloni \& Hasinger \shortcite{belloni:90a} wherein integrating over
Fourier frequency yields the mean square variability relative to the square
of the mean of the lightcurve.

All three PSDs had roughly comparable shapes and amplitudes.  Specifically,
they show evidence for a rise in power at frequencies $\aproxlt 3 \times
10^{-3}$\,Hz, broad peaks at $f \sim 0.02$ and $0.08$\,Hz, and reasonably
sharp declines for $f \aproxgt 0.1$\,Hz. No power in excess of noise is
seen above $\approx 0.3$\,Hz.  The root mean square (rms) variability
increases from soft to hard energies. From softest to hardest energy bands,
the rms variabilities are 8\%/7\%, 10\%/8\%, and 11\%/9\% over the
$10^{-3}$--0.3\,Hz/$10^{-2}$--0.3\,Hz range. The up-turn at low frequency
is most likely due to variability associated with the orbital time scales.
(Due to low count rates, phase resolved variability studies are very
difficult.)  The coherence at $f \aproxlt 10^{-2}$\,Hz being slightly lower
than that at higher frequencies might be related to an admixture of
intrinsic variability of the corona with variability associated with the
orbital period, although overall the varibiality between the highest and
lowest energy bands is well-correlated. We note that 7--9\% rms variability
in the $10^{-2}$--0.3\,Hz range is consistent with the variability seen in
both low ($\aproxlt 5\%~L_{\rm Edd}$) and high ($\aproxgt 30\%~L_{\rm
  Edd}$) fractional Eddington luminosity neutron star and black hole
sources.

The lack of high frequency variability may be due to scattering over large
distances and/or optical depths (see Nowak \& Vaughan \nocite{nowak:96a}
1996, and references therein). Specifically, if the variable lightcurve is
first passed through a scattering medium of optical depth $\tau$ and size
$D$, one expects a cutoff in the power spectrum at a frequency, $f_{\rm
  cut}$, given by
\begin{equation}
2 \pi f_{\rm cut} \approx {\rm min}\left [ \frac{c}{\tau D}, \frac{c}{D}
     \right ] ~~.
\label{eq:cut}
\end{equation}
Thus, given $D \sim R_c \sim 0.2a \sim 3 \times 10^{10}$\,cm, the cutoff
frequency should be $f_{\rm cut} \approx 0.2~{\rm
  min}[\tau^{-1},1]$\,Hz\footnote{For an input spectrum confined to a
  narrow energy band scattered into a narrowly observed output energy band,
  the cutoff frequency can be given by $2 \pi f_{\rm cut} \approx {c}/D$,
  even for $\tau \aproxgt 1$.  This is because for such a situation, fixing
  the energy band of the observed output is essentially fixing the number
  of scatters that the observed photons have undergone, and therefore
  narrows the dispersion in the photon arrival times.  See Nowak \& Vaughan
  \shortcite{nowak:96a}, and references therein.}, consistent with what is
observed here.

Scattering may also lead to the observed variability lags.  Over the narrow
frequency region for which one can obtain lag measurements, the softest
energy band is seen to lag the hardest energy band by $\approx 1$\,s.
(Note, the statistics were not sufficient to measure accurately the time
lag of the variability in the middle energy band with respect to that in
either of the other two energy bands.)  Such a lag is consistent with a
central X-ray source being reprocessed downward in energy by scattering
within a corona of radius/scattering path length of oder $3\times
10^{10}~{\rm cm}$, consistent with the fits of \S\ref{sec:lcfit}.

\section{Discussion}\label{sec:discuss}
Having laid out the observational facts, we now turn to their
interpretation.  Several key issues about \eight\ have surfaced in the
earlier sections, which we will now address, along with some open questions
which relate to earlier studies of this object.

\subsection{Detailed Spectral Fitting}
As we have shown, several fundamentally different models can explain the
\asca\ and \rxte\ spectra almost equally well. While formally satisfactory,
the spectra we presented in \S\ref{sec:spectra} still show some systematic
residues, which could be removed by addition of features like absorption
edges and lines.  For example, P00 argue that there is clear evidence for
an edge around 1.3\,keV.  We do see a hint of such an effect, especially in
the optically thick case, but it is not clear if the origin is really an
absorption edge. The same feature might be produced by an emission line
complex at 1 keV (as demonstrated in the optically thin case, where the
addition of a Raymond-Smith plasma removed the systematic residues at the
low energy end). We also note that the edge in the fits by P00 falls very
close to the position where two major spectral components (blackbody
and Comptonization spectrum) cross in their fits.

Similarly, the structure of the iron line region is all but clear. As
remarked by P00, the ratio of the two narrow lines we fit to the data is
off from the theoretical value expected for K${\alpha}$ to K${\beta}$, but
ionized line emission could explain this fact. In the optically thin models
the ionization parameter in the corona is much larger than one, which would
strengthen such an argument. Both of these questions will undoubtedly be
answered by upcoming \xmm\ and \axaf\ observations.

\subsection{Optically Thick vs. Thin}
Both in spectral modelling and in fitting the light curve, we used two
general models, based on whether the corona is optically thick or thin.  We
found that both light curve and spectrum are degenerate with respect to
this distinction, and optically thick and thin models reproduce these data
equally well.  We will now discuss the physical interpretation of the two
cases (see also \S\ref{sec:wind}).

\begin{itemize}
\item{In the optically thick case we modelled the spectrum as a partially
    absorbed power law, produced by reprocessed radiation from the corona,
    passing through a column of cold gas (the atmosphere of the outer
    disc).  From the spectral fit we can then deduce the covering fraction
    and optical depth of the partial absorber. These numbers are given in
    Table \ref{tab:thick}.
    
    We can understand the above mentioned correlation between the partial
    absorption column, the absorption fraction, and the overall flux
    through the following picture: the X-rays originate from an
    approximately spherical corona (at uniform surface brightness).  The
    accretion disc rim is completely opaque and responsible for most of the
    modulation in the light curve, due to differences in its height (and
    thus its covering fraction).  Above the accretion disc, an atmosphere
    of cold gas absorbs/scatters a fraction of the light from the corona
    behind it. The geometric covering fraction would roughly be given by
    the ratio of the atmosphere's scale height to the coronal radius.  As
    the rim height changes through the binary phases, the covering fraction
    of the partial absorber also changes (see Table~\ref{tab:thick}).
    
    Higher latitudes of the corona are relatively less absorbed, thus for
    binary phases where the the disc rim is higher, the ratio of unabsorbed
    to absorbed coronal flux will be greater.  At the same time, the average
    column depth of the absorber should be smaller, since the gas density
    is lower at higher latitudes.  The trend in Fig.~\ref{fig:depths} can
    also be understood this way: the absorbed component of the power law
    contributes primarily hard X-rays to the spectrum.  Since this part of
    the spectrum originates at low latitudes, it is most affected by
    obscuration, thus the larger modulation depths at high energies.  The
    implied scale heights and column depths of Table \ref{tab:thick} are
    consistent with what would be expected from an X-ray heated atmosphere
    above the disc between the edge of the corona and the disc rim
    \cite{white:82a}.
    
    The fact that we only see a small fraction of the luminosity inferred
    from the orbital evolution can be explained if the corona is optically
    thick to scattering while relatively optically thin to absorption.  In
    that case it will act as a mirror and transmit only a fraction
    $(1+\tau)^{-1}$ of the incident radiation.  For a ratio of observed to
    inferred luminosity of $\sim 200$, modulo a geometric covering factor,
    this would imply an optical depth of $\tau \aproxlt 200$ (which
    justifies the assumption of an optically thick corona in this model).
    As we will discuss below (see \S\ref{sec:wind}), a large fraction of
    the energy could also leave the system in a wind (Adiabatic
    Inflow-Outflow Solution, ADIOS; Blandford \& Begelman
    \nocite{blandford:99a} 1999) or, in the case of a black hole primary,
    disappear down the horizon (optically thick Advection Dominated
    Accretion Flow, ADAF; Narayan \& Yi \nocite{narayan:94a} 1994).
    
    Table \ref{tab:thick} shows that the equivalent widths of the iron
    lines are rather large, of order 250 eV.  A simple estimate of the
    expected EW produced solely within the partial absorber by the incident
    coronal spectrum is of the order of 50 eV. Uncertainties in the
    metallicity or the geometry of the system (i.e., if the partial
    absorber is subject to a larger flux than seen by the observer) might
    account for such a discrepancy. Note also that the partial covering
    model introduces an iron absorption edge at 7.1 keV, which is leading
    in part to the large fitted equivalent widths.  As noted by P00 and
    references therein, the line ratios are still problematic if the higher
    energy line is an Fe K$\beta$ line, as opposed to ionized Fe K$\alpha$.
    Models of resonant scattering of the Fe K$\alpha$ line by the partial
    absorber in the outer disc rim region may help explain some of this
    discrepancy. Finally, we note that the optically thick corona itself
    could well be the origin of strong iron line emission.}
  
\item{The key question for the optically thin case is whether or not there
    is unambiguous evidence for blackbody emission from the surface of the
    compact object that is scattered into our line of sight.  Related to
    this is the question of whether or not there is clear evidence for
    coronal line emission, i.e. a thermal plasma component.  As the
    optically thick model, with partial absorption, fits the spectrum
    adequately, evidence for these components cannot be claimed to be
    unambiguous.  Ignoring the partial covering (due to the limits of our
    data and models), however, these components are seen to be allowed in
    the fits, thus there absence cannot be definitively shown either.
    
    Assuming the optically thin scenario to be correct, the X-ray emission
    is again produced in the corona. The blackbody part of the continuum is
    produced either on the neutron star surface (if the central object is
    in fact a neutron star) or the inner disc, and is then scattered into
    our line of sight by the corona. The power law part of the spectrum
    stems from radiation that is Comptonized in the corona. If the corona
    is optically thin, the fraction of the total X-ray luminosity that
    reaches the observer is $\approx \tau$.  The optical depth of the
    corona would thus be $\tau \aproxgt 1/200$, once again modulo the
    geometric covering factor. We note again that a fraction of the
    accretion energy might be carried away in a wind or advected (in the
    case of a black hole primary).
    
    Note that the modulation seen in the power law is stronger than the
    modulation seen in the blackbody.  We can interpret this dependence
    geometrically if we further postulate that the power law emission is
    produced at lower coronal latitudes, in which case a larger fraction of
    it would be subject to obscuration both by the rim and the companion.
    The variations shown in Fig. \ref{fig:depths} would be natural in such
    a scenario, since the black body component emanates from the whole
    corona, is relatively less modulated, and contributes most strongly at
    low and intermediate energies.  Again, however, a partial covering
    model may be applicable for the optically thin case as well as for the
    optically thick case.
    
    The fitted iron line in this case is weaker than for the optically
    thick case, but due to the neglect of partial absorption this model
    does not contain an intrinsic absorption feature around 7 keV. All
    other remarks with respect to the line strength in the optically thick
    case hold here too.}
\end{itemize}
Both these models can explain the basic features of light curve and
spectrum, we therefore have no way to chose one over the other.  In fact,
reality might fall in between these two cases.  One might imagine a model
where the optical depth of the corona decreases with height.  We would thus
see scattered blackbody radiation from high latitudes and a partially
absorbed power law from the optically thick parts closer to the disc.
Hopefully, future X-ray and optical observations will resolve this
degeneracy.

\subsection{NS, BH, or WD?}
Lacking detections of X-ray pulses or Type I X-ray bursts that would
unambiguously point toward a neutron star primary, we must rely on more
indirect arguments for determining its nature. As discussed in
\S\ref{sec:lcfit}, the fits to the X-ray lightcurves are ambiguous in this
regard.  Of the possibilities, we find the white dwarf primary scenario the
least likely for two principle reasons.  First is the fact that we observe
an X-ray flux of order $10^{36}~{\rm ergs~s^{-1}}$.  Eq.~\ref{eq:mdot},
assuming no mass loss and a more typical white dwarf radiative efficiency
of $\eta \approx 10^{-3}$, would lead to a total accretion luminosity of
$\approx 4 \times 10^{35}~{\rm ergs~s^{-1}}$. Although by assuming a white
dwarf primary we decrease the inferred size of the disc and thereby
decrease the inferred distance \cite{mason:82a} and luminosity by a factor
of $\sim 0.6$ and $\sim 0.4$, respectively, we would still require that we
are viewing of order 100\% of the accretion luminosity despite the presence
of a scattering corona and the near edge-on inclination of the system.
Second is the implied mass of the secondary, $M_2 \sim 0.06~\msun$, which
is slightly small given the then inferred Roche lobe radius of the
secondary, $R_2 \approx 2 \times 10^{10}~{\rm cm} \approx 0.3~{\rm
  R}_\odot$. Again we note, however, that if the lower values of
$K_1=70$\,km~s$^{-1}$ and $K_2=225$\,km~s$^{-1}$ are correct
\cite{harlaftis:97a}, than a white dwarf primary is the preferred model.

Of the oter two possibilities, neutron star or black hole, there is little
to distinguish between them.  Given the edge-on inclination and the implied
large scattering paths, it is not surprising that we do not detect any
variability associated with a neutron star spin period, as discussed in
\S\ref{sec:var}.  Furthermore, given little or no mass loss, the implied
accretion rate onto the neutron star would be sufficient to suppress any
Type I bursting behaviour (Bildsten \nocite{bildsten:95a} 1995, and
references therein). We therefore expect a neutron star and low mass black
hole to look nearly identical. Distinguishing between these two scenarios,
therefore, will require a combination of more careful observations of the
secondary and identifying a plausible evolutionary scenario for this
system.

\subsection{Winds, Advection Domination, and the Evolutionary History of
  \eight}
\label{sec:wind}

As discussed in \S\ref{sec:orbit}, the implied mass transfer time scale is
$\tau_{\rm mt} \sim 10^{7}{\rm years}$, which is compatible with mass
transfer on a thermal time scale (see Kalogera \& Webbink
\nocite{kalogera:96a} 1996, and references therein).  One interesting
possibility is that we are viewing the \eight\ system towards the end of a
thermal time scale mass transfer phase \cite{kalogera:96a}.  Specific
scenarios have been discussed for Cyg~X-2 \cite{king:99a,king:99b} and
SS~433 \cite{king:00a}. At the onset of mass transfer the system begins
with a mass ratio $q > 1$.  As the secondary loses mass, both the binary
separation and the secondary Roche lobe shrink.  A common envelope phase is
avoided by strong mass loss from the primary, perhaps in an advection
dominated accretion phase characterized by a strong wind emanating from
large radii \cite{king:99b,king:00a}.  Even as $q$ decreases below unity,
mass transfer can continue to be driven on the thermal time scale. For the
Cyg~X-2 system (known to be a neutron star primary; Smale
\nocite{smale:98a} 1998), it is hypothesised that the system began with an
$\approx 3.5~\msun$ secondary, but now shows an $\approx 0.5~\msun$
secondary with a radius $\approx 7~{\rm R}_\odot$ \cite{king:99a}.

The above mentioned systems, however, cannot be exactly analogous to that
of the \eight\ system. Cyg~X-2, for example, has a 9.84\,day orbital period
and a secondary luminosity of $\approx 150~{\rm L}_\odot$, both far larger
than in \eight. In the optical bands, aside from X-ray heating of the
secondary, it has traditionally been assumed that the \eight\ emission is
dominated by the accretion disc and disc rim, with the total optical
luminosity being $L_{\rm opt} \sim 10~{\rm L}_\odot$ \cite{mason:82a}.
Likewise, models of the UV emission have been presumed to be dominated by
the disc and the corona \cite{mason:82c}, although perhaps the far UV
allows the greatest room for significant contributions from the secondary.
In addition, obtaining the currently observed 5.57\,hr orbital period via
mass transfer alone requires that during the earlier epoch wherein $0.2
\aproxlt q \aproxlt 1$, the orbital period was even shorter.  The current
5.57\,hr orbital period might require an extended epoch of magnetic braking
of the secondary (Rappaport et al.  \nocite{rappaport:83a} 1983; Taam
\nocite{taam:83a} 1983; Pylyser \& Savonije \nocite{pylyser:88a} 1988; and
references therein).

As discussed in the above references, magnetic braking is expected only to
be effective for secondaries that do not have a fully convective envelope.
Although strong mass transfer rates are possible before mass loss leads to
such an envelope, this epoch is only expected to last $10^6$--$10^7$ years
\cite{rappaport:83a}.  This is consistent with our previous estimates of
the lifetime of the \eight\ system.  We note that during this strong mass
transfer phase, one generally expects the secondary to exceed its main
sequence radius \cite{taam:83a,kalogera:96a}.  If we define the ratio of
the secondary's Roche lobe radius to its main sequence radius as ${\cal
  R}$, then given the discussion of \S\ref{sec:lcfit} we can approximate
the mass of the primary as
\begin{equation}
M_1 \approx 3.0~ {\cal R}^{-9/4} \left ( \frac{K_1/\sin i}{70~{\rm
      km~s^{-1}}/\sin 85^\circ} \right )^{-2/3} ~~.
\end{equation}
Thus, ${\cal R} \aproxgt 1.2$ allows for a  neutron star mass $\aproxlt
2~\msun$. 

Finally we note that in addition to any mass loss associated with magnetic
braking, further strong mass loss from such a system as suggested by King
\& Begelman \shortcite{king:99a} leads to another explanation (in addition
to postulating that the coronal optical depth $\tau \ll 1$ or $\tau \gg 1$)
for the fact that we observe only $\approx 10^{-2}$ of the accretion
luminosity inferred for an efficiency of $\eta \approx 10\%$.  `ADIOS'
models wherein most of the accreted mass never reaches the surface or event
horizon of the primary \cite{blandford:99a,king:99a} allow for either a
neutron star or black hole primary.  In such a scenario, the energy of
viscous dissipation is not efficiently radiated away from the system and
must be carried away via a wind. Viscous dissipation occurs out to the
circularization radius, which is the inferred radius of the corona from our
fits of \S\ref{sec:lcfit}, and thus dissipation may in part be responsible
for the presence of the corona (which in this case could be an ADIOS wind).

\section{Conclusions}\label{sec:conclusions}

We have presented observations of the low mass X-ray binary \eight\, taken
with both \rxte\ and \asca. We considered two broad band fits, which we
took as approximately representing `optically thick' and `optically thin'
coronal emission.  Either model fit the data nearly equally well.
Likewise, the X-ray lightcurves folded on the orbital period were also
equally well-fit by optically thick or thin models, and furthermore these
fits could not distinguish among a white dwarf, neutron star, or black hole
primary. High spectral resolution \asca\ data revealed complex structure in
the Fe K$\alpha$/K$\beta$ region, consisting of possibly two lines.  These
latter features might be related to obscuring material between the edge of
the corona and the disc rim, and also possibly related to emission from any
optically thick regions of the corona.

The spectral ambiguities of the line region will likely be resolved by
upcoming high resolution observations with \xmm\ and \axaf. The nature of
the primary, however, will likely not be revealed by these observations.
As discussed above, understanding the nature of the primary will in large
part depend upon further observations of the secondary, and depend upon
identifying a plausible evolutionary scenario for the \eight\ 
system.\smallskip

This research has been supported NSF Grants AST95--29170 and AST98--76887
(SH) and NASA grant NAG5-3225 (MAN). We would like to thank N. White for
allowing us to use his \asca\ and \rxte\ data.  We would also like to
acknowledge useful conversations with D.  Chakrabarty, J.  Chiang, R.
Moderski, C.  Reynolds, R. Taam, J.  Wilms, and E.  Zweibel.  A debt of
gratitude is owed to P. Maloney for sharing with us his own special genius.
This research has made use of data obtained through the High Energy
Astrophysics Science Archive Research Center Online Service, provided by
the NASA/Goddard Space Flight Center.

\appendix

\section{Data Analysis Methodology}
\subsection{RXTE Data Analysis}\label{sec:rxte_anal}

We extracted data from both pointed instruments on \rxte, the Proportional
Counter Array, \pca, and the High Energy X-ray Timing Experiment, \hexte.
\eight, however, is both very faint and very soft, therefore we only
spectrally fit data from the \pca\ instrument.  The \rxte\ data were
analyzed using the same procedure as that for our analysis of the spectrum
of GX~339$-$4 \cite{wilms:99a}.  Specifically, all \rxte\ results in this
paper were obtained using the standard \rxte\ data analysis software, {\tt
  ftools} version 4.2, and response matrix v3.1.  Data selection criteria
were that the source elevation was larger than 10$^\circ$ above the earth
limb and data measured within 30\,minutes of passages of the South Atlantic
Anomaly or during times of high particle background (as expressed by the
``electron ratio'' being greater than 0.1) were ignored.  To increase the
signal to noise level of the data, we restricted the analysis to the first
anode layer of the proportional counter units (PCUs) where most source
photons are detected (the particle background is almost independent of the
anode layer), and we combined the data from all five PCUs.  We only used
data where all five PCUs were turned on, which was nearly the entire
observation.

For spectral fitting, we limited the energy range of the \pca\ data from 3
to 30\,keV To take into account the calibration uncertainty of the \pca\ we
applied the channel dependent systematic uncertainties described by Wilms
et al.\,\shortcite{wilms:99a}.  These uncertainties were determined from a
power-law fit to an observation of the Crab nebula and pulsar taking into
account all anode chains; however, they do also provide a good estimate for
the first anode layer only since most of the photons are detected in this
layer. Background subtraction of the \pca\ data was performed using the
`SkyVLE' model, as for our previous studies of GX~339$-$4
\cite{wilms:99aa}.

\subsection{ASCA Data Extraction}\label{sec:asca_anal}

We extracted data from the two solid state detectors (\sis 0, \sis 1) and
the two gas detectors (\gis 2, \gis 3) onboard \asca\ by using the standard
ftools as described in the ASCA Data Reduction Guide \cite{day:98a}.  We
chose circular extraction regions with radii of $\approx 4$\,arcmin for the
SIS detectors, and $\approx 6$\,arcmin for the \gis\ detectors.  We
excluded approximately the central 1\,arcmin to avoid the possibility of
photon pileup.  We used the {\tt sisclean} and {\tt gisclean} tools (with
default values) to remove hot and flickering pixels. We filtered the data
with the strict cleaning criteria outlined by Brandt et al.
\shortcite{brandt:96a}; however, we took the larger value of
7\,$\mbox{GeV}/c$ for the rigidity.  We rebinned the spectral files so that
each energy bin contained a minimum of 20\,photons. We retained \sis\ data
in the 0.5 to 10\,keV range and \gis\ data in the 1 to 10\,keV range. The
background was measured from rectangular regions on the two edges of the
chip farthest from the source (\sis\ data), or from annuli with inner radii
$>6$\,arcmin (\gis\ data).  These data were cleaned and filtered in the
same manner as the source files.

Note that for the simultaneous \asca/\rxte\ observations, we combined the
two \sis\ detectors into a single spectrum, and we combined the two \gis\
detectors into a single spectrum, properly weighting the response matrices.
We accounted for the cross-calibration uncertainties of the \sis\ and \gis\
instruments relative to each other and relative to \rxte\ by introducing a
multiplicative constant for each detector in all of our fits.  Note also
that the phase filtering option in {\tt xselect} produces flawed results,
so the time filtering option was used instead to produce phase selected
spectra.


\end{document}